\documentstyle[epsf]{elsart}
\begin{document}
\begin{frontmatter}
\title{Nonlinear Debye-Onsager-Relaxation-Effect}
\author{K. Morawetz}
\address{Fachbereich Physik, Universit\"at Rostock,
18051 Rostock, Germany}

\begin{abstract}
The quantum kinetic equation for charged particles in strong electric fields 
is used to derive the nonlinear particle flux. The relaxation field is calculated quantum mechanically up to any order in the applied field provided a given Maxwellian plasma. The classical limit is given in analytical form. In the range of weak fields the deformation of the screening cloud is responsible for the Debye-Onsager relaxation effect. 
\end{abstract}
\end{frontmatter}

\begin{figure}
\epsfxsize=15cm
\centerline{\epsffile{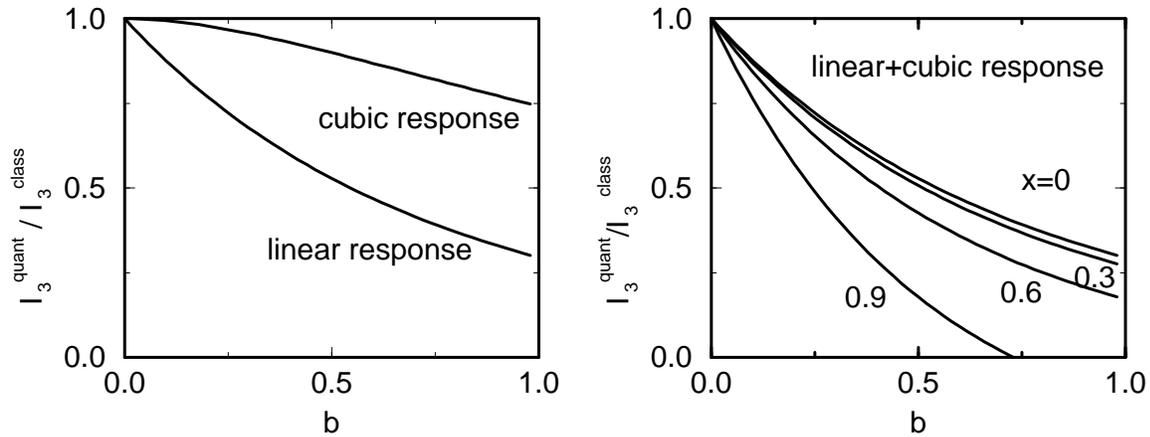}}
\vspace{2ex}
\caption{\label{2}The ratio of quantum to classical 
Debye-Onsager relaxation effect (\protect\ref{form}) versus quantum parameter $b$. In the left hand 
figure the linear (\protect\ref{sum}) and cubic response term in the expansion of $x$ is plotted separately. In the 
right hand figure we give the relaxation effect up to cubic terms for different field strength $x$.}
\end{figure}

High field transport has become a topic of current interest in various fields of physics. Especially, the nonlinear current or the electrical conductivity gives access to properties of dense nonideal plasmas. 
At low fields we expect the linear response regime to be valid. Then the contribution of field effects to the conductivity can be condensed into the Debye- Onsager relaxation effect \cite{k58,KE72,er79,r88,MK92} which was first derived within the theory of electrolytes \cite{DH23,O27,f53}. Debye has given a limiting law of electrical conductivity which stated that the external electric field $E$ on a charge $e$ is diminished by the amount
$\delta E=E \,(1- {\kappa e^2 \over 6 \epsilon_0 T})
$
 where $\kappa$ is the inverse screening radius of the screening cloud. Later it has been shown by Onsager that this result has to be corrected if 
the dynamics of ions is considered.
In this paper we will give the complete result beyond linear response for the static case
$E \,(1- {\kappa e^2 \over 6 \epsilon_0 T} F(E))
$
similar to the theoretical explanation of the Wien effect \cite{f53}.
We start from the kinetic equation derived with the help of the gauge invariant formulation of Green's function, 
\protect\cite{JW84,Moa93}
\begin{eqnarray}\label{kinetic}
&&\frac{ \partial}{\partial t}  f_a + e {\bf E} {\nabla_{\bf k_a}} f_a =
\sum_b  I_{ab} \nonumber \\
\vspace*{2cm}
I_{ab} &=&
\frac{2}{\hbar^2}\int \frac{ d {\bf k'_a} d {\bf k_b} d {\bf k'_b}
}{(2\pi\hbar)^6}V_{ab}^2({\bf k_a}- {\bf k'_a})
\delta \left({\bf k_a}+{\bf k_b}- {\bf k'_a}-{\bf k'_b} \right)\nonumber\\
&\times&
\int\limits_0^{\infty} d\tau
\cos
\left\{
(\epsilon_a+\epsilon_b-{\epsilon'_a}-{\epsilon'_b}){\tau \over \hbar} \right .
- \left .
\frac{{\bf E}\tau^2}{2 \hbar}
\left(
\frac{e_a{\bf k_a}}{m_a}+ \frac{e_b{\bf k_b}}{m_b}-
\frac{e_a {\bf k'_a}}{m_a}-\frac{e_b {\bf k'_b}}{m_b}
\right)
\right\}\nonumber\\
&\times&\left\{
f'_a f'_b(1-{f}_a)(1-{f}_b)-{f}_a {f}_b (1-f'_a)(1-f'_b)
\right\}.
\end{eqnarray}
Here we have written, e.g., $f_b$ for $f_b(k_b-e_bE\tau,t-\tau)$
for simplicity. We use the static screened Debye potential here, which restriction can be released to dynamical screening \cite{MO97}.
Generalizations can be found for the T-matrix \cite{Mor94} approximation resulting into a field dependent Bethe-Salpeter 
equation or for the RPA approximation \cite{Moa93} resulting into a Lenard-Balescu kinetic equation.
We are now interested in corrections
to the particle flux, and therefore obtain from (\ref{kinetic}) the balance equation for the momentum
\begin{equation}\label{p}
\frac{\partial}{\partial t} <{\bf k_a}>-ne_a{\bf E} =
\sum_b<{\bf k_a}I_B^{ab}>\equiv n_ae_a{\bf E}{\delta
E_a\over E}.
\end{equation}
We assume a nondegenerate
situation, such that the Pauli blocking effects can be neglected and assume a 
quasistationary plasma with Maxwellian distributions, which in principle restricts the applied fields \cite{MK93}.
The angular
integrations can be carried out trivially and we get
\begin{eqnarray}\label{sum}
&&<{\bf p}I_B^{ab}>={{\bf E} \over E} \sum\limits_b{8 n_a n_b e_a^2 e_b^2 \over \tilde T \epsilon_0^2} I_3\nonumber\\
I_3&=&2 \sum\limits_{k=0}^\infty (-1)^{k+1}{k+1 \over (2 k+3)!}
x^{2 k+1} I_3^{2 k+1}\nonumber\\
I_3^{2 k+1}&=&\int\limits_0^{\infty} dz {z^{2 k+4} \over (z^2+1)^2}
\int\limits_0^{\infty} dl \, l^{4 k+2} \,{\sin(z^2 l b ) \over b}
{\rm e}^{-z^2 l^2}\nonumber\\
&=&{(2 k+1)! \over 2}\int\limits_0^{\infty} dz {z^{2 -2 k} \over (z^2+1)^2}
\, {\rm _1 F_1}(2+2 k, \frac 3 2, -{b^2 z^2 \over 4})
\end{eqnarray}
with $\tilde T=\frac 1 2 \left ({m_b \over m_a+m_b} T_a+{m_a \over m_a+m_b}
T_b \right )$ and the quantum parameter
$b^2={\hbar^2 \kappa^2 \over 4 \mu \tilde T}
$
and the classical field parameter
$x={E \over 2 \tilde T \kappa}\left ({m_a \over m_a+m_b}
e_b-{m_b \over m_a+m_b}
e_a \right ).$
First we give here an exact expression for the classical limit.
We observe that (\ref{sum}) for $b\rightarrow 0$ diverges.
However,  we can calculated the classical limit directly
\begin{eqnarray}
I_{3c}&=&-{\pi x \over 24} F(|x|)\nonumber\\
F^{\rm static}(x)&=&-\frac{3}{x^2}\left [ 3- x +{1 \over 1+x} -\frac 4 x {\rm
ln}(1+x) \right ]=1+o(x)\nonumber\\
F^{\rm dyn}(x)&=&-\frac{3}{x^2}\left [ 2- x +{2 \over x}{\rm
ln}(1+x) \right ]=2+o(x).\label{classical}
\end{eqnarray}
The second result we calculated for inclusion of dynamical screening within the approximation \cite{er79} which replaces $\overline{\epsilon(\omega,q)^{-2}}$ by $(1+\kappa^2V_{aa}(q)/4\pi)^{-1}$.
This result gives the 
classical field dependent Debye- Onsager- relaxation- effect up
to field strengths $x<1$.
Introducing the classical result (\ref{classical}) into (\ref{p}) the following relaxation field appears
\begin{eqnarray}\label{form}
{\delta E_a \over E}&=&{e_a \pi \over 12 \kappa} \sum\limits_b
{4 n_b e_b^2\over \epsilon_0^2 \mu}{{e_b\over m_b}-{e_a\over
m_a}\over \left ({T_b\over m_b}+{T_a\over
m_a}\right )^2} F\left ({E\over \kappa} {{e_b\over m_b}-{e_a\over m_a}\over {T_a\over m_a}+{T_b\over m_b}}\right ).
\end{eqnarray}
No relaxation field appears for particles with equal charge to
mass ratios. The link to the known
Debye- Onsager relaxation effect can be found if we assume that we
have a plasma consisting of heavy ion ($a$) with charge one and oppositely charge light ions ($b$) and
temperatures $T_a=T_b=T$. Then (\ref{form}) reduces to
${\delta E_a\over E}=-{\kappa e_a^2\over 12 \epsilon_0 T} F({e E \over T \kappa})$
and $F(x)$ from (\ref{classical}).
Within the linear response the dynamical result leads to the known Debye relaxation field \cite{MO97} while the static result here underestimates the value about one half. The high field result $F(x)$ is monotonously approaching zero for high fields and can be compared with the known result from electrolyte theory, recently \cite{O97}. The result (\ref{form}) with (\ref{classical}) is an extension to the work of \cite{MK92} in that it gives the relaxation field up to any field strength, not restricting to linear response and an extension to \cite{Ma97} that dynamical screening is included.
The complete quantum case of (\protect\ref{sum}) can be given by performing the integral. The result 
gives a series in field strength $x$, which however does not converge for $x=1$. 
In the following we give only the first two parts of the 
expansion with respect to the field. The quantum effects are included completely. 
The quantum linear response reads
\begin{eqnarray}
&&I_3^1(k=0)={\pi \over 8}  \,\left( 1 + {b_1^2} - 
       b_1 \,{\sqrt{\pi }} {{\rm e}^{{{{b_1^2}}}}} \left ( \frac 3 2 +{b_1^2} \right )
        {\rm erfc}({b_1})  \right)
\end{eqnarray}
with $b_1=b/2$.
This result reproduces \cite{MK92} by a different way of 
calculation. 
All higher order terms can be given in analytical form as well.
In the Fig. \ref{2}a we plot the quantum versus classical result for 
linear response and cubic response versus the quantum parameter $b$. We see that the cubic response is less 
influenced by quantum effects than the linear response result. The general observation is that the quantum effects lower 
the classical result for the relaxation effect. A detailed analysis of quantum effects on the linear response can be found 
in \cite{MK92,Ma97}. 

In Fig. \ref{2}b we give the ratio of quantum to classical result for the relaxation effect up to cubic terms in fields for 
different field strengths x. We see that the quantum effects become more important 
with increasing field strength. The effect of sign change can be seen in 
the quantum effects at certain values of $b$.
We remark that the electric field is limited to values $x<1$ or 
$E<{\kappa T\over e}
$
 beyond no quasi equilibrated transport is possible, i.e. no thermal distributions are 
pertained in the system. Then we have to take into account nonthermal field dependent distributions which have been 
employed to study nonlinear conductivity \cite{KMSR92,MK93}.

This work was supported from the BMBF (Germany) under contract
Nr. 06R0884.

\end{document}